\documentclass[prd,twocolumn,showpacs,floatfix,amsmath,nofootinbib,amssymb,floatfix]{revtex4}
\usepackage{graphicx,color,dcolumn,booktabs,bm}
\usepackage{longtable,lscape}
\usepackage{pdfpages}
\usepackage{txfonts}
\usepackage{overpic}
\usepackage{amssymb}
\usepackage{indentfirst}
\usepackage{feynmf}   
\usepackage{slashed}  
\usepackage{cases}
\usepackage{color}
\usepackage{multirow}
\usepackage{threeparttable}
\usepackage{epstopdf}
\usepackage{enumerate}
\usepackage{graphicx,color,dcolumn,booktabs,bm}
\usepackage[colorlinks, citecolor=blue,anchorcolor=red,menucolor=red, linkcolor=red,filecolor=red,urlcolor=blue,frenchlinks=red]{hyperref}

\graphicspath{{Figures/}} %

\begin{document}

\title{Interpretation of the observed $\Lambda_b(6146)^0$ and $\Lambda_b(6152)^0$ states as 1$D$ bottom baryons}
\author{Bing Chen$^{1,3}$}\email{chenbing@ahstu.edu.cn}
\author{Si-Qiang Luo$^{2,3}$}\email{luosq15@lzu.edu.cn}
\author{Xiang Liu$^{2,3}$\footnote{Corresponding author}}\email{xiangliu@lzu.edu.cn}
\author{Takayuki Matsuki$^{4,5}$}\email{matsuki@tokyo-kasei.ac.jp}
\affiliation{$^1$School of Electrical and Electronic Engineering, Anhui Science and Technology University, Fengyang 233100, China
\\$^2$School of Physical Science and Technology, Lanzhou University, Lanzhou 730000, China\\
$^3$Research Center for Hadron and CSR Physics, Lanzhou University $\&$ Institute of Modern Physics of CAS, Lanzhou 730000, China
\\$^4$Tokyo Kasei University, 1-18-1 Kaga, Itabashi, Tokyo 173-8602, Japan
\\$^5$Theoretical Research Division, Nishina Center, RIKEN, Wako, Saitama 351-0198, Japan}

\date{\today}

\begin{abstract}

The measured masses and strong decays of $\Lambda_b(6146)^0$ and $\Lambda_b(6152)^0$ required us to decode them as the 1$D$ excited states with $J^P=3/2^+$ and $J^P=5/2^+$, respectively. Under the suggested assignment, the masses and total decay widths of $\Lambda_b(6146)^0$ and $\Lambda_b(6152)^0$ can be explained. As a $J^P=3/2^+$ state, the $\Lambda_b(6146)^0$ should mainly decay into $\Sigma_b(5815)\pi$. However, this theoretical result is in contradiction to the measurement since no significant $\Lambda_b(6146)^0\rightarrow\Sigma_b(5815)^\pm\pi^\mp$ signals were observed by the LHCb. The possible explanations to this difficulty are mentioned. For completeness, we present the prediction of masses and decay widths of other unseen 1$D$ bottom baryons in this work. Our results could provide some important clues for the upcoming experiments.


\end{abstract}
 \maketitle

\section{Introduction}\label{sec1}
Single heavy baryons occupy an important position in the hadron's jigsaw puzzle since chiral symmetry and heavy quark symmetry (HQS) can provide some qualitative insight into their dynamics. So the investigation of single heavy baryons, including the charm and bottom baryons, could be more helpful for improving our understanding of the confinement mechanism. With the experimental and theoretical efforts, the spectrum of both charm and bottom baryon families is being established step by step~\cite{Tanabashi:2018oca}. Here, we briefly review some important measurements of the excited charm and bottom baryons by the different experiments in the past years.

In the $\Lambda_c^+$ sector, a new resonance, denoted as $\Lambda_c(2860)^+$, was discovered by the LHCb Collaboration~\cite{Aaij:2017vbw}. The spin-parity of $\Lambda_c(2860)^+$ was identified as $3/2^+$. In the same work, the LHCb also confirmed the spin of $\Lambda_c(2880)^+$, which was first measured by the Belle Collaboration~\cite{Abe:2006rz}. Specifically, the preferred spin of the $\Lambda_c(2880)^+$ state was found to be $J=5/2$. In addition, the first constraint on the spin-parity of the $\Lambda_c(2940)^+$ state was performed in the same analysis~\cite{Aaij:2017vbw}, where the $\Lambda_c(2940)^+$ was most likely a $J^P=3/2^-$ state. Belle recently determined the isospin of $\Lambda_c(2760)^+/\Sigma_c(2760)^+$ and disentangled the identity of this puzzling state~\cite{Abdesselam:2019bfp}. Since no evidence of $\Sigma_c(2760)^{++,0}$ was observed in the $\Sigma_c(2455)^{++/0}\pi^0$ channels, there is now no doubt that the puzzling state around 2.76 GeV is a $\Lambda_c^+$ state.

Experiments have also made a progress in the study of excited $\Xi_c$ and $\Omega_c$ states. The $\Xi_c(3055)^+$ and $\Xi_c(3080)^+$ were first observed in the decay channel of $\Lambda{D^+}$, and their relevant ratios of branching fractions were also measured~\cite{Kato:2016hca}. Five narrow $\Omega_c$ states, denoted as $\Omega_c(3000)$, $\Omega_c(3050)$, $\Omega_c(3065)$, $\Omega_c(3090)$, and $\Omega_c(3120)$, were found by the LHCb Collaboration~\cite{Aaij:2017nav} and were subsequently confirmed by the Belle Collaboration~\cite{Yelton:2017qxg}.


For the excited bottom baryons, the 1$P$ $\Lambda_b^0$ states, i.e., $\Lambda_b^0(5912)^0$ and $\Lambda_b^0(5920)^0$, were established by LHCb~\cite{Aaij:2012da} and CDF~\cite{Aaltonen:2013tta}. Last year, two bottom baryons, denoted as $\Xi_b(6227)^-$~\cite{Aaij:2018yqz} and $\Sigma_b(6097)^{\pm}$~\cite{Aaij:2018tnn}, were first observed by the LHCb Collaboration. Very recently, the LHCb Collaboration again announced their new discovery of two bottom baryon states, the $\Lambda_b(6146)^0$ and $\Lambda_b(6152)^0$, by analyzing the $\Lambda_b^0\pi^+\pi^-$ invariant mass spectrum from $pp$ collisions~\cite{Aaij:2019amv}. The resonance parameters of these two states are listed below,
\begin{equation}
\begin{aligned}\label{eq1}
m_{\Lambda_b(6146)^0}=~&6146.17\pm0.33\pm0.22\pm0.16 \textrm{~MeV},\\
m_{\Lambda_b(6152)^0}=~&6152.51\pm0.26\pm0.22\pm0.16 \textrm{~MeV},\\
\Gamma_{\Lambda_b(6146)^0}=~&2.9\pm1.3\pm0.3 \textrm{~MeV},\\
\Gamma_{\Lambda_b(6152)^0}=~&2.1\pm0.8\pm0.3 \textrm{~MeV}.
\end{aligned}
\end{equation}

These newly observed single heavy baryons have motivated a wide discussion and study of their inner structures~\cite{Chen:2016spr}. In the recent years, our group has also carried a systematic study of the low-lying charm and bottom baryons, mainly including the 2$S$ and 1$P$ states in the single heavy baryon family~\cite{Chen:2014nyo,Chen:2016iyi,Chen:2017aqm,Chen:2017gnu,Chen:2018orb,Chen:2018vuc}. In this work, we will further investigate the probable assignment of the new $\Lambda_b(6146)^0$ and $\Lambda_b(6152)^0$ and try to establish the 1$D$ bottom baryons.

Heavy quark symmetry can provide us with some qualitative insight of the $\Lambda_b(6146)^0$ and $\Lambda_b(6152)^0$ when we compare them with the established $\Lambda_c^+$ baryon states. At present, the 1$S$, 1$P$, and 1$D$ $\Lambda_c^+$ baryons have been established~\cite{Tanabashi:2018oca}. As shown later, the excited energies of $\Lambda_b(6146)^0$ and $\Lambda_b(6152)^0$ indicate that the 1$D$ assignment is suitable for these two new bottom baryons. The predicted masses and strong decays can further test this assignment.

With the coming LHCb Upgrade I in 2020~\cite{Bediaga:2018lhg}, more and more excited bottom baryons are expected to be discovered in the near future. So it is necessary for us to systematically investigate the properties of the whole $D$-wave bottom baryons by including their mass spectrum and strong decay behaviors, which can provide some valuable information for further experiments to explore the excited bottom baryons.

The paper is organized as follows. In Sec. \ref{sec2}, a brief review of our theoretical methods will be presented, where the parameters will also be given for the reader's convenience. We decode the newly reported $\Lambda_b(6146)^0$ and $\Lambda_b(6152)^0$ as the $D$-wave states in Sec. \ref{sec3}, where the analysis of the mass spectrum and the investigation of strong decays will be presented. For completeness, we will also predict the mass spectra and strong decays of other $D$-wave bottom baryons in Sec. \ref{sec4}. Finally, the paper ends with a discussion and conclusions in Sec. \ref{sec5}.

\section{Scheme of the quark potential model and the quark pair creation (QPC) model}\label{sec2}

Heavy quark symmetry plays an important role in the dynamics of a single heavy baryon system. In the heavy quark limit, one heavy quark within the heavy baryon system is decoupled from two light quarks. Although heavy quark symmetry is broken due to the finite charm or bottom quark mass, much evidence indicates that heavy quark symmetry is a good approximation for the single heavy baryons.

\begin{figure}[htbp]
\begin{center}
\includegraphics[width=3.2cm,keepaspectratio]{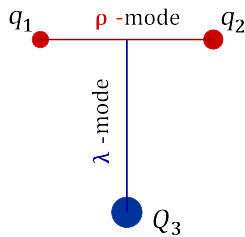}
\caption{The so-called ``$\rho$-mode'' and ``$\lambda$-mode'' excitations in a single heavy baryon system. The quarks $q_1$ and $q_2$ represent the light $u$, $d$, and $s$ quarks, while the $Q_3$ quark refers to a heavy charm or bottom quark.}\label{Fig1}
\end{center}
\end{figure}

Under the scenario of heavy quark symmetry, the dynamics of a heavy baryon state could be well separated into two parts~\cite{Yoshida:2015tia}. As illustrated in Fig. \ref{Fig1}, the degree of freedom between two light quarks ($q_1$ and $q_2$) is usually called ``$\rho$-mode,'' while the degree between the center of mass of two light quarks and the heavy quark is called ``$\lambda$-mode.'' If adopting the ordinary confining potential like the linear or harmonic form to depict the interaction of quarks, the excited energy of the $\rho$-mode is higher than that of the $\lambda$-mode excitations~\cite{Copley:1979wj,Yoshida:2015tia}. So it seems to be easier for the experiments to detect the $\lambda$-mode excited heavy baryons. The available theoretical investigations confirm the point that almost all observed single heavy baryons could be explained as the $\lambda$-mode excited states.

In Refs.~\cite{Chen:2018orb,Chen:2018vuc}, we assigned the $\Xi_b(6227)^-$ and $\Sigma_b(6097)^{\pm}$ as the $P$-wave $\lambda$-mode excited bottom baryons. In this work, we also examine the possibility of $\Lambda_b(6146)^0$ and $\Lambda_b(6152)^0$ as the $\lambda$-mode excitations. Since two light quarks in a bottom baryon are always in the ground state with the antitriplet color structure, the $\lambda$-mode bottom heavy baryon can be simplified as a quasi-two-body system, which means that the diquark could be an effective degree of freedom in these $\lambda$-mode excited bottom baryons. To depict the interaction between the light diquark and the bottom quark, the following Schr\"{o}dinger equation can be constructed,
\begin{equation}
\left(-\frac{\nabla^2}{2m_\mu}-\frac{4\alpha}{3r}+br+C+\frac{32\alpha\sigma^3}{9\sqrt{\pi}m_{di}m_b}\vec{\textrm{s}}_{di}\cdot\vec{\textrm{s}}_{b} \right)\psi_{nL} = E\psi_{nL}. \label{eq2}
\end{equation}
Here, $\vec{\textrm{s}}_{di}$ and $\vec{\textrm{s}}_{b}$ denote the spin of the light diquark and the $b$ quark, respectively. The reduced mass is defined as $m_\mu~{\equiv}~m_{di}m_b/(m_{di}+m_b)$. The parameters $\alpha$, $b$, and $C$ stand for the strength of the color Coulomb potential, the strength of linear confinement, and a mass-renormalized constant, respectively. The parameter $\sigma$ comes from a Gaussian-smeared contact hyperfine interaction between the light diquark and the bottom quark. Since the spin of the light diquark in the $\Lambda_b$ and $\Xi_b$ baryons is zero, the spin-spin contact hyperfine interaction in Eq.~(\ref{eq2}) is only important to calculate the radially excited $\Sigma_b$, $\Xi_b^\prime$, and $\Omega_b$ states. The values of the parameters which have been given in Ref.~\cite{Chen:2018orb} are also listed in Table~\ref{table1} for the reader's convenience.

\begin{table}[htbp]
\caption{Values of the parameters for the bottom bayons in the nonrelativistic quark potential model where the mass of the $b$ quark is taken as 4.96 GeV. Here, $m_{di}$ refers to the mass of different diquarks.
}\label{table1}
\renewcommand\arraystretch{1.1}
\begin{tabular*}{86mm}{c@{\extracolsep{\fill}}ccccc}
\toprule[1pt]\toprule[1pt]
 Parameters       & $m_{di}$ (GeV)    & $\alpha$       & $b$  (GeV$^2$)   & $\sigma$ (GeV)    & $C$ (GeV)  \\
\toprule[1pt]
$\Lambda_b$       & 0.45              & 0.20           & 0.112            & $-$               & 0.265  \\
$\Xi_b$           & 0.63              & 0.26           & 0.118            & $-$               & 0.176 \\
$\Sigma_b$        & 0.66              & 0.22           & 0.116            & 1.20              & 0.185  \\
$\Xi^{\prime}_b$  & 0.78              & 0.22           & 0.116            & 1.20              & 0.152  \\
$\Omega_b$        & 0.91              & 0.26           & 0.120            & 1.07              & 0.120  \\
\bottomrule[1pt]\bottomrule[1pt]
\end{tabular*}
\end{table}
By solving the Schr\"{o}dinger equation, the spin-averaged masses of these excited bottom baryons can be obtained. When the spin-orbit and tensor interactions are incorporated, all masses of $D$-wave $\lambda$-mode excited bottom baryons can be calculated. Furthermore, the simple harmonic oscillator (SHO) wave function $\psi_{nL}^m=\mathcal{R}_{nL}(\beta,\textbf{K})\mathcal{Y}_{nL}^m(\textbf{K})$ is usually adopted to construct the spatial wave function of the bottom baryon states and other hadron states when one uses the QPC model to calculate the strong decays of excited bottom baryons. The SHO wave-function scale $\beta$ is also fixed by solving the above Schr\"{o}dinger equation. In other word, the parameters in the potential model and in the QPC decay model have the same values except for the unique parameter $\gamma$ of
the QPC model.

As mentioned above, we will use the QPC model to study the strong decays of $D$-wave excited bottom baryons. For an OZI-allowed decay process of a hadron state, the QPC model assumes that a $q\bar{q}$ pair is created from the vacuum and then regroups into two outgoing hadrons by a quark rearrangement process~\cite{Micu:1968mk,LeYaouanc:1972vsx,LeYaouanc:1988fx}. For a process ``$A\rightarrow{B+C}$,''  the transition matrix element in the QPC model is written as $\langle{BC}|\mathcal {\hat{T}}|A\rangle=\delta^3(\textbf{\textrm{K}}_B+\textbf{\textrm{K}}_C)\mathcal {M}^{j_A,j_B,j_C}(p)$ where the transition operator $\mathcal {\hat{T}}$ reads as
\begin{equation}
\begin{split}
\mathcal {\hat{T}}=&-3\gamma
\sum_{\text{\emph{m}}}\langle1,m;1,-m|0,0\rangle \iint
d^3\textbf{\textrm{k}}_{\mu}d^3\textbf{\textrm{k}}_{\nu}\delta^3(\textbf{\textrm{k}}_{\mu}+\textbf{\textrm{k}}_{\nu})\\ &\times\mathcal
{Y}_1^m\left(\frac{\textbf{\textrm{k}}_{\mu}-\textbf{\textrm{k}}_{\nu}}{2}\right)\omega_0^{({\mu},{\nu})}\varphi^{({\mu},{\nu})}_0\chi^{({\mu},{\nu})}_{1,-m}d^\dag_{{\mu}}(\textbf{\textrm{k}}_{\mu})d^\dag_{{\nu}}(\textbf{\textrm{k}}_{\nu}),
\end{split}\label{eq3}
\end{equation}
in a nonrelativistic limit. Here, the dimensionless parameter $\gamma$ describes the strength of the quark-antiquark pair created from the vacuum. The $\omega_0^{({\mu},{\nu})}$ and $\varphi^{({\mu},{\nu})}_0$ are the color and flavor wave functions of the $q_{\mu}\bar{q}_{\nu}$ pair created from the vacuum. Therefore, $\omega_0^{({\mu},{\nu})}=(R\bar{R}+G\bar{G}+B\bar{B})/\sqrt{3}$ and $\varphi^{({\mu},{\nu})}_0=(u\bar{u}+d\bar{d}+s\bar{s})/\sqrt{3}$ are color and flavor singlets. The $\chi^{({\mu},{\nu})}_{1,-m}$ represents the pair production in a spin-triplet state. When the mock state~\cite{Hayne:1981zy} is adopted to describe the wave function of a hadron state, the partial wave amplitudes $\mathcal{M}_{LS}(p)$ can be obtained by the following formula,
\begin{equation}
\begin{aligned}\label{eq4}
\mathcal {M}^{A\rightarrow B+C}_{LS}(p)=&\frac{\sqrt{2L+1}}{2J_A+1}\sum_{\text{$j_B$,$j_C$}}\langle L0Sj_A|J_Aj_A\rangle\\
&\times\langle J_Bj_B,J_Cj_C|Sj_A\rangle\mathcal
{M}^{j_A,j_B,j_C}(p).
\end{aligned}
\end{equation}
Here, $p$ represents the momentum of an outgoing meson in the rest frame of a meson $A$. The $J_i$ and $j_i$ ($i=$ $A$, $B$, and $C$) denote the total angular momentum and the projection of initial and final hadron states, respectively, and $L$ denotes the orbital angular momenta between the final state $B$ and $C$. Finally, the partial width of $A\rightarrow BC$ can be obtained in terms of the partial wave amplitudes
\begin{equation}
\begin{aligned}\label{eq5}
\Gamma(A\rightarrow BC)=2\pi\frac{E_BE_C}{M_A}p\sum_{L,S}|\mathcal
{M}^{A\rightarrow B+C}_{LS}(p)|^2,
\end{aligned}
\end{equation}
in the $A$ rest frame. Interested readers can consult Refs.~\cite{Chen:2016iyi,Chen:2017gnu} for more details of our method.

\section{Decoding $\Lambda_b(6146)^0$ and $\Lambda_b(6152)^0$ as the $D$-wave excited states}\label{sec3}

The masses of 1$D$ $\Lambda_b^0$ baryons have been calculated in our previous works~\cite{Chen:2014nyo,Chen:2018vuc} and in Refs.~\cite{Capstick:1986bm,Roberts:2007ni,Ebert:2011kk}. The results in Refs.~\cite{Chen:2014nyo,Chen:2018vuc} are in good agreement with the measured masses of $\Lambda_b(6146)^0$ and $\Lambda_b(6152)^0$, while the predicted masses from Refs.~\cite{Capstick:1986bm,Roberts:2007ni,Ebert:2011kk} are about 30$-$40 MeV larger than the corresponding experimental results.\footnote{The mass of the ground $\Lambda_b^0$ state was given as 5585 MeV in Ref.~\cite{Capstick:1986bm}, which is about 35 MeV below the $\Lambda_b(5620)^0$ state. If one shifts up the predicted mass to match the measured mass of $\Lambda_b(5620)^0$, then the predicted masses of 1$P$ and 1$D$ $\Lambda_b^0$ states in Ref.~\cite{Capstick:1986bm} are also about 30$-$45 MeV larger than the measurements~\cite{Aaij:2012da,Aaltonen:2013tta,Aaij:2019amv}.} Here, we should point out that the authors in Refs.~\cite{Capstick:l1986bm,Roberts:2007ni,Ebert:2011kk} adopted the same parameters to predict the masses of whole heavy baryon states including the bottom and charm baryons. In fact, the input parameters of the quark potential model could be slightly different for the charm and bottom baryons. And, when considering the uncertainties of the quark potential model, the results presented in Refs.~\cite{Capstick:1986bm,Roberts:2007ni,Ebert:2011kk} do not contradict the 1$D$ assignment to $\Lambda_b(6146)^0$ and $\Lambda_b(6152)^0$.

In this work, we also give a semiquantitative analysis to illustrate the possibility of $\Lambda_b(6146)^0$ and $\Lambda_b(6152)^0$ as the 1$D$ $\Lambda_c^+$ states. With the observations of $\Lambda_c(2286)^+$, $\Lambda_c(2595)^+$, $\Lambda_c(2625)^+$, $\Lambda_c(2860)^+$, and $\Lambda_c(2880)^+$ states, the 1$S$, 1$P$, and 1$D$ $\Lambda_c^+$ baryons have been established by experiments~\cite{Tanabashi:2018oca}. In the $\Lambda_b^0$ sector, the 1$S$ and 1$P$ states have also been established~\cite{Tanabashi:2018oca}. By comparing with the excited energies of $\Lambda_c(2860)^+$ and $\Lambda_c(2880)^+$, it is possible to further decode the $\Lambda_b(6146)^0$ and $\Lambda_b(6152)^0$ as the 1$D$ excited states.

\begin{figure}[htbp]
\begin{center}
\includegraphics[width=8.6cm,keepaspectratio]{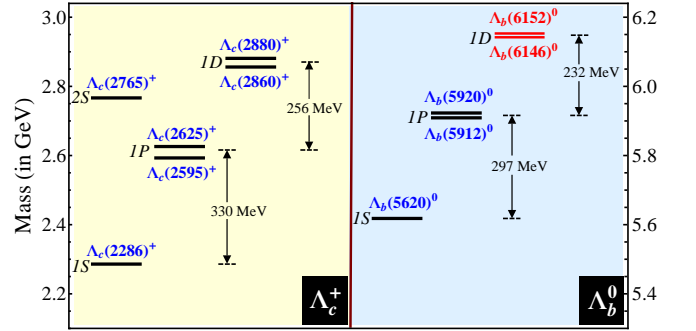}
\caption{The established $\Lambda_c^+$ and $\Lambda_b^0$ states and the newly observed $\Lambda_b(6146)^0$ and $\Lambda_b(6152)^0$.}\label{Fig2}
\end{center}
\end{figure}

The mass gaps between the 1$P$/1$S$ and 1$D$/1$P$ states are illustrated in Fig.~\ref{Fig2} for the $\Lambda_c^+$ and $\Lambda_b^0$ baryons. The mass gap between the 1$P$ and 1$S$ $\Lambda_c^+$ states, which we denote as $\Delta_{1P-1S}(\Lambda_c^+)$, is about 330 MeV. The gap between the 1$D$ and 1$P$ $\Lambda_c^+$ states is about 256 MeV, i.e., $\Delta_{1D-1P}(\Lambda_c^+)\approx256$ MeV. Then, the ratio
\begin{equation}
\begin{aligned}\label{eq6}
\varepsilon(\Lambda_c^+)=\frac{\Delta_{1D-1P}(\Lambda_c^+)}{\Delta_{1P-1S}(\Lambda_c^+)}=\frac{256~\textrm{MeV}}{330~\textrm{MeV}}=0.776,
\end{aligned}
\end{equation}
is obtained. Similarly, the ratio of $\Lambda_b^0$ states is given as
\begin{equation}
\begin{aligned}\label{eq7}
\varepsilon(\Lambda_b^0)=\frac{\Delta_{1D-1P}(\Lambda_b^0)}{\Delta_{1P-1S}(\Lambda_b^0)}=\frac{232~\textrm{MeV}}{297~\textrm{MeV}}=0.781.
\end{aligned}
\end{equation}
Obviously, the ratios of $\Delta_{1D-1P}/\Delta_{1P-1S}$, which are nearly equal for the $\Lambda_c^+$ and $\Lambda_b^0$ baryons, imply the similarity of dynamics between the charm and bottom baryons. The result of $\varepsilon(\Lambda_c^+)\approx\varepsilon(\Lambda_b^0)$ can be partly explained by the potential model. For simplicity, we replace the Cornell potential in Eq. (\ref{eq2}) by the following confining potential,
\begin{equation}
\begin{aligned}\label{eq8}
V(r)=\kappa{r}^\nu,
\end{aligned}
\end{equation}
to depict the interaction between the light diquark and heavy quark ($c$ or $b$ quark) in the single heavy baryon system. In this way, the orbitally excited energy can be obtained analytically as
\begin{equation}
\begin{aligned}\label{eq9}
E_L=\kappa^{\frac{2}{\nu+2}}\left(\frac{1}{2m_\mu}\right)^{\frac{\nu}{\nu+2}}\left[\sqrt{\pi}\nu\left(L+\frac{3}{2}\right)\frac{\Gamma\left(\frac{1}{\nu}+\frac{3}{2}\right)}{\Gamma\left(\frac{1}{\nu}\right)}\right]^{\frac{2\nu}{\nu+2}},
\end{aligned}
\end{equation}
by the Wentzel-Kramers-Brillouin (WKB) approach. Then the ratio defined by Eqs.~(\ref{eq6}) and (\ref{eq7}) can be given by
\begin{equation}
\begin{aligned}\label{eq10}
\varepsilon=\frac{5^{\frac{2\nu}{\nu+2}}-7^{\frac{2\nu}{\nu+2}}}{3^{\frac{2\nu}{\nu+2}}-5^{\frac{2\nu}{\nu+2}}},
\end{aligned}
\end{equation}
which is independent of the parameter $\kappa$ and the reduced mass $m_\mu$. When the parameter $\nu$ is taken as $0.8\pm0.2$, the $\varepsilon$ is predicted as $0.835\pm0.035$. 

According to the above analysis, the new $\Lambda_b(6146)^0$ and $\Lambda_b(6152)^0$ states could be treated as the partners of the $\Lambda_c(2860)^+$ and $\Lambda_c(2880)^+$ in the bottom baryon family. Furthermore, the mass splitting of $\Lambda_b(6146)^0$ and $\Lambda_b(6152)^0$, which was measured as $\Delta{m}=6.34\pm0.32\pm0.02$ MeV, is also consistent with the expectations of $1D$ $\Lambda_b^0$ states~\cite{Chen:2014nyo,Ebert:2011kk,Chen:2017fcs,Chen:2016phw}. So according to the measured masses, the $\Lambda_b(6146)^0$ and $\Lambda_b(6152)^0$ states could be good candidates for $D$-wave bottom baryons.

Thus, it is desirable to further test the $D$-wave assignment of $\Lambda_b(6146)^0$ and $\Lambda_b(6152)^0$ by exploring their strong decays. A method based on the QPC model (see Sec. \ref{sec2}) has successfully explained the strong decays of known bottom baryons~\cite{Chen:2018orb,Chen:2018vuc} including the new $\Xi_b(6227)^-$ and $\Sigma_b(6097)^{\pm}$ states. With the same method and parameters, we calculate the strong decays of the $\Lambda_b(6146)^0$ and $\Lambda_b(6152)^0$ and list the results in Table~\ref{table2}.

\begin{table}[htbp]
\caption{Partial widths of strong decays of $\Lambda_b(6146)^0$ and $\Lambda_b(6152)^0$ as the $D$-wave excited states (in MeV). The superscript letters $p$ and $f$ mean that the corresponding decays occur via the $p$-wave and $f$-wave, respectively.}\label{table2}
\renewcommand\arraystretch{1.3}
\begin{tabular*}{86mm}{c@{\extracolsep{\fill}}cc}
\toprule[1pt]\toprule[1pt]
 Decay mode                 & $\Lambda_b(6146)^0$ $[3/2^+~(1D)]$     & $\Lambda_b(6152)^0$ $[5/2^+~(1D)]$   \\
\toprule[1pt]
$\Sigma_b(5815)~\pi$        & 3.25$^p$                               & 0.22$^f$             \\
$\Sigma^\ast_b(5835)~\pi$   & 0.65$^p$,~~~0.28$^f$                   & 4.03$^p$,~~~0.14$^f$              \\
Total width                 & 4.18                                   & 4.39              \\
\toprule[1pt]
Expt.~\cite{Aaij:2019amv}   & 2.9~$\pm$~1.3~$\pm$~0.3                & 2.1~$\pm$~0.8~$\pm$~0.3              \\
\bottomrule[1pt]\bottomrule[1pt]
\end{tabular*}
\end{table}

The total widths of $\Lambda_b(6146)^0$ and $\Lambda_b(6152)^0$ are obtained as 4.18 MeV and 4.39 MeV when they are assigned as the $D$-wave states with $J^P=3/2^+$ and $J^P=5/2^+$, respectively. The predicted total widths of $\Lambda_b(6146)^0$ and $\Lambda_b(6152)^0$ are comparable with the experimental measurements~\cite{Aaij:2019amv}. However, the LHCb announced that they did not observe significant $\Lambda_b(6146)^0\rightarrow\Sigma_b(5815)^\pm\pi^\mp$ signals in their measurements~\cite{Aaij:2019amv}. This experimental result seems to contradict the theoretical results in Table~\ref{table2}. The plausible explanations of this contradiction are given as follows.

As almost degenerate states, the $\Lambda_b(6146)^0$ and $\Lambda_b(6152)^0$ have the same decay channels (see Table~\ref{table2}). Then, the possible interference effects which were ignored in the LHCb's analysis~\cite{Aaij:2019amv} might be important to pin down the decay behaviors of $\Lambda_b(6146)^0$ and $\Lambda_b(6152)^0$. Another possible explanation is that the mass inversion may occur for the 1$D$ $\Lambda_b^0$ states~\cite{Liang:2019aag,Wang:2019uaj}. That is to say, the $\Lambda_b(6146)^0$ with smaller mass should be the $J^P=5/2^+$ state, while the $\Lambda_b(6152)^0$ should be a $J^P=3/2^+$ state. However, this assignment of $\Lambda_b(6146)^0$ and $\Lambda_b(6152)^0$ disagrees with the expectations given by the quark potential models~\cite{Capstick:1986bm,Roberts:2007ni,Ebert:2011kk}. So more experimental and theoretical efforts are required for the $\Lambda_b(6146)^0$ and $\Lambda_b(6152)^0$ states in the future.


\section{Other predicted 1$D$ bottom baryons}\label{sec4}


The basis $|{s}_{di},L_{\lambda},j_{di},s_b,J\rangle$ including heavy quark symmetry will be used in our scheme. Here, ${s}_{di}$ and $s_b$ denote the spins of the light diquark and bottom quark, respectively; $L_{\lambda}$ denotes the orbital quantum number between the light diquark and the bottom quark. Then, we define  $\vec{j}_{di}=\vec{s}_{di}+\vec{L}_{\lambda}$ and $\vec{J}=\vec{s}_b+\vec{j}_{di}$. For the $D$-wave $\Lambda_b^0$ and $\Xi_b^{0,-}$ baryons, $j_{di}=L_{\lambda}=2$ since $s_{di}=0$. For the $D$-wave $\Sigma_b^{0,\pm}$, $\Xi_b^{\prime0,-}$, and $\Omega_b^-$ baryons, $j_{di}=1\otimes2=1$, 2, and 3 with $s_{di}=1$. For simplicity, we denote these bottom baryons as $|{J}^P,1D\rangle_{j_{di}}$. Finally, we list the abbreviated basis in Table \ref{table3}.

\begin{table}[htbp]
\caption{The abbreviated basis of $D$-wave bottom baryons.}\label{table3}
\renewcommand\arraystretch{1.3}
\begin{tabular*}{86mm}{c@{\extracolsep{\fill}}lll}
\toprule[1pt]\toprule[1pt]
    Bottom baryons                  & \multicolumn{3}{c}{Basis}     \\
\toprule[1pt]
$\Lambda_b^0$, $\Xi_b^{0,-}$                            & $\mid3/2^+,1D\rangle_2~$        & $\mid5/2^+,1D\rangle_2~$      &~       \\
$\Sigma_b^{0,\pm}$, $\Xi_b^{\prime0,-}$, $\Omega_b^-$   &  $\mid1/2^+,1D\rangle_1$     & $\mid3/2^+,1D\rangle_1$    & $\mid3/2^+,1D\rangle_2$            \\
                                                        &  $\mid5/2^+,1D\rangle_2$     & $\mid5/2^+,1D\rangle_3$    &  $\mid7/2^+,1D\rangle_3$          \\
\bottomrule[1pt]\bottomrule[1pt]
\end{tabular*}
\end{table}

In practice, the mass of the $b$ quark is not infinite. Then, two physical baryons with the same spin-parity should be the mixing states of $|{J}^P,1D\rangle_{J-1/2}$ and $|{J}^P,1D\rangle_{J+1/2}$. We take the $D$-wave $\Sigma_b$ baryons with $J^P=3/2^+$ as an example. In principle, two physical $3/2^+$ $\Sigma_b$ baryons should be the mixing states of $|3/2^+,1D\rangle_1$ and $|3/2^+,1D\rangle_2$,
\begin{eqnarray}
\begin{aligned}
 \left(
           \begin{array}{c}
                     |3/2^+,1D\rangle_H\\
                     |3/2^+,1D\rangle_L\\
                    \end{array}
     \right)&=\left(
           \begin{array}{cc}
                    \cos\theta   & -\sin\theta  \\
                    \sin\theta   & ~~\cos\theta  \\
                    \end{array}
     \right)  \left(
           \begin{array}{c}
                     |3/2^+,1D\rangle_1 \\
                     |3/2^+,1D\rangle_2 \\
                    \end{array}
     \right). \label{eq11}
\end{aligned}
\end{eqnarray}
Here, the physical $J^P=3/2^+$ states are distinguished by their different masses. Specifically, the states with the lower and higher masses are denoted by the subscripts ``$L$'' and ``$H$,'' respectively. Our practical calculations indicate that the mixing effect is quite small for these $D$-wave bottom states with the same $J^P$. For simplicity, we could ignore these small mixing effects and denote the physical states by the $|{J}^P,1D\rangle_{j_{di}}$ basis directly.

\subsection{$1D$ $\Xi_b$ baryons}

Due to the approximate $SU(3)$ flavor symmetry, the $\Lambda_b$ and $\Xi_b$ states may have similar properties. With the observation of $\Lambda_b(6146)^0$ and $\Lambda_b(6152)^0$, the 1$S$, 1$P$, and 1$D$ $\Lambda_b$ states may have been established by experiments. Differently from $\Lambda_b$, only the ground states, $\Xi_b(5792)^0/\Xi_b(5797)^-$, have been observed in the $\Xi_b$ sector~\cite{Tanabashi:2018oca}. With the running of LHCb,  we may expect the excited $\Xi_b$ baryons to be observed in the near future. The masses and decay properties of the $1P$ and $2S$ $\Xi_b$ states have been investigated in our previous work~\cite{Chen:2018orb}. In the following, we give the predicted properties, especially, the strong decay behaviors of the $D$-wave $\Xi_b$ baryons.

The masses of two $D$-wave $\Xi_b$ baryons with $J^P=3/2^+$ and $J^P=5/2^+$ were predicted as 6327 MeV and 6330 MeV, respectively (see Fig. 2 in Ref.~\cite{Chen:2018orb}). With the predicted masses, the partial and total decay widths obtained by the QPC model are listed in Table~\ref{table4}.

\begin{table}[htbp]
\caption{Partial widths of strong decays of the $D$-wave $\Xi_b$ baryon states (in MeV). The superscript letters $p$ and $f$ represent the corresponding decays occuring via the $p$-wave and the $f$-wave, respectively.}\label{table4}
\renewcommand\arraystretch{1.3}
\begin{tabular*}{86mm}{c@{\extracolsep{\fill}}cc}
\toprule[1pt]\toprule[1pt]
 Decay mode                   & $\Xi_b(6327)$ $[3/2^+~(1D)]$     & $\Xi_b(6330)^0$ $[5/2^+~(1D)]$   \\
\toprule[1pt]
$\Xi^{\prime}_b(5935)~\pi$    & 0.39$^p$                               & 0.09$^f$             \\
$\Sigma_b(5815)~K$            & 1.73$^p$                               & 0.00$^f$             \\
$\Xi^\ast_b(5955)~\pi$        & 0.09$^p$,~~~0.15$^f$                   & 0.51$^p$,~~~0.07$^f$              \\
$\Sigma^\ast_b(5835)~K$       & 0.02$^p$,~~~0.00$^f$                   & 0.09$^p$,~~~0.00$^f$              \\
Total width                   & 2.38                                   & 0.76              \\
\bottomrule[1pt]\bottomrule[1pt]
\end{tabular*}
\end{table}

According to the results, the $D$-wave $\Xi_b$ baryons seem to be the narrow resonances. It is possible to find the $3/2^+(1D)$ $\Xi_b$ baryon in the $\Xi^{\prime}_b(5935)~\pi$ and $\Sigma_b(5815)~K$ channels, while the $5/2^+(1D)$ $\Xi_b$ baryon might be found in the decay channel of $\Xi^\ast_b(5955)~\pi$. A similar conclusion was also obtained by the chiral quark model~\cite{Yao:2018jmc}.

\subsection{$1D$ $\Sigma_b$ baryons}

The masses of 1$D$ $\Sigma_b$ baryons have been predicted in our previous works~\cite{Chen:2018orb,Chen:2018vuc} (see Fig. 1 in Ref.~\cite{Chen:2018vuc}). For clarity, we list the predicted masses of 1$D$ $\Sigma_b$ and all other bottom baryons together in Table \ref{table5}. According to the predictions, six $D$-wave $\Sigma_b$ baryons may exist in the region 6.29$-$6.40 GeV. Then many decay channels are open for these high excited $\Sigma_b$ baryons.

\begin{table}[htbp]
\caption{Predicted masses of 1$D$ $\Lambda_b$, $\Xi_b,$ $\Sigma_b$,  $\Xi^\prime_b$, and $\Omega_b$ baryons (in MeV).
}\label{table5}
\renewcommand\arraystretch{1.3}
\begin{tabular*}{86mm}{c@{\extracolsep{\fill}}cccccc}
\toprule[1pt]\toprule[1pt]
 States         & $|1/2^+\rangle_1$  & $|3/2^+\rangle_1$  & $|3/2^+\rangle_2$  & $|5/2^+\rangle_2$  & $|5/2^+\rangle_3$  & $|7/2^+\rangle_3$    \\
\toprule[1pt]
$\Lambda_b$     &                       &                       & 6149                  & 6153                  &                       &                       \\
$\Xi_b$         &                       &                       & 6327                  & 6330                  &                       &                       \\
$\Sigma_b$      & 6400                  & 6402                  & 6358                  & 6359                  & 6291                  & 6292                  \\
$\Xi_b^\prime$  & 6486                  & 6488                  & 6456                  & 6457                  & 6407                  & 6408                  \\
$\Omega_b$      & 6599                  & 6602                  & 6578                  & 6580                  & 6542                  & 6543                  \\
\bottomrule[1pt]\bottomrule[1pt]
\end{tabular*}
\end{table}

The predicted partial and total widths of the $D$-wave $\Sigma_b$ baryons are presented in Table \ref{table6}. In our calculation of strong decays, if the final states have been found, we take the measured masses as inputs. Otherwise, we take the predicted masses of the unknown final states as inputs. As shown in Table \ref{table6}, the $|1/2^+\rangle_1$, $|3/2^+\rangle_1$, $|3/2^+\rangle_2$, and $|5/2^+\rangle_2$ $\Sigma_b$ states might be broad since their decay widths are predicted around 100 MeV. So it is a challenge to search for them in experiments. Differently from these states, the $|5/2^+\rangle_3$ and $|7/2^+\rangle_3$ $\Sigma_b$ states are much narrower. Experimentalists may search these two states in the $\Lambda_b(5620)\pi$, $\Lambda_b(5912)\pi$, and $\Lambda_b(5920)\pi$ channels.

\begin{figure}[htbp]
\begin{center}
\includegraphics[width=5.2cm,keepaspectratio]{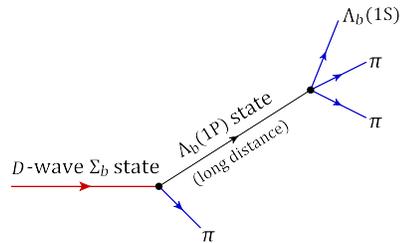}
\caption{The diagram to illustrate the $D$-wave $\Sigma_b$ baryon decaying into a $P$-wave $\Lambda_b$ baryon and a $\pi$ meson. Subsequently, the $P$-wave $\Lambda_b$ baryon decays into the ground $\Lambda_b$ baryon and two $\pi$ mesons.}\label{Fig3}
\end{center}
\end{figure}

\begin{table*}[htbp]
\caption{The predicted partial and total widths of $D$-wave $\Sigma_b$ baryons (in MeV). The superscript letters $s$, $p$, $\cdots$, and $h$ mean that the corresponding decays occur via $s$-, $p$-, $\cdots$, and $h$-waves, respectively. If the partial width is predicted to be smaller than 0.01 MeV, it is denoted as ``0.00.'' If a decay channel is forbidden, it is denoted by ``$\times$.'' If the mass of an initial state is below the threshold of a decay channel, the partial width is denoted by ``$-$.'' The masses of 1$P$ and 2$S$ $\Sigma_b$ baryons are borrowed from our previous work~\cite{Chen:2018vuc}.} \label{table6}
\renewcommand\arraystretch{1.3}
\begin{tabular*}{170mm}{l@{\extracolsep{\fill}}cccccc}
\toprule[1pt]\toprule[1pt]
Final states    &    $\Sigma_b(6400)$         & $\Sigma_b(6402)$         &    $\Sigma_b(6358)$      &    $\Sigma_b(6359)$      &    $\Sigma_b(6291)$   &     $\Sigma_b(6292)$     \\
\midrule[1pt]
 $\Lambda_b(5620)~\pi$        & 1.89$^p$  & 1.88$^p$               &   $\times$            &  $\times$            &    7.14$^f$       &   7.13$^f$      \\
 $\Xi_b(5795)~K$              & 2.16$^p$  & 2.16$^p$               &   $\times$            &  $\times$            &    0.00$^f$       &   0.00$^f$      \\
 $\Sigma_b(5815)~\pi$         & 0.27$^p$  & 0.06$^p$               &   1.47$^p$            &  2.05$^f$            &    1.64$^f$       &   0.93$^f$      \\
 $\Sigma^\ast_b(5835)~\pi$    & 0.17$^p$  & 0.41$^p$               & 0.37$^p$,~~~3.96$^f$  & 2.17$^p$,~~~1.77$^f$ &    2.08$^f$       &   2.83$^f$      \\
 $\Lambda_b(5912)~\pi$        & 12.35$^s$ & 1.49$^d$               &   10.63$^d$           &  4.73$^d$            &    8.72$^d$       &   3.17$^g$      \\
 $\Lambda_b(5920)~\pi$        & 2.18$^d$  & 12.58$^s$,~~~1.17$^d$  &   10.53$^d$           &  16.41$^d$           & 2.51$^d$,~~~4.62$^g$ & 11.27$^d$,~~~1.95$^g$ \\
 $\Sigma_b(6150)~\pi$         &  $\times$ &  $\times$              &   0.39$^d$            &  0.40$^d$            &    $\times$       &   $\times$        \\
 $\Sigma_b(6134)~\pi$         & 13.91$^s$ & 1.62$^d$               &   1.74$^d$            &  0.79$^d$            &    0.01$^d$       &    0.00$^g$           \\
 $\Sigma_b(6139)~\pi$         & 2.74$^d$  & 14.93$^s$,~~~1.43$^d$  &   1.50$^d$            &  2.39$^d$            & 0.00$^{d,~g}$     & 0.01$^d$,~~~0.00$^g$  \\
 $\Sigma_b(6094)~\pi$         & 8.45$^d$  & 4.30$^d$               & 26.41$^s$,~~~1.16$^d$ & 0.34$^d$,~~~0.15$^g$ & 1.18$^d$,~~~0.00$^g$ & 0.15$^d$,~~~0.00$^g$   \\
 $\Sigma_b(6098)~\pi$         & 5.36$^d$  & 9.67$^d$               & 0.47$^d$,~~~0.18$^g$  & 27.80$^s$,~~1.26$^d$,~~0.06$^g$  & 0.24$^d$,~~~0.00$^g$  & 1.15$^d$,~~~0.00$^g$  \\
 $\Lambda_b(6086)~\pi$        & 13.05$^p$ & 13.30$^p$              &   $\times$            &  $\times$            &    0.06$^f$       &   0.07$^f$      \\
 $\Lambda_b(6146)~\pi$        & 15.88$^p$ & 1.57$^p$,~~~0.04$^f$   & 11.00$^p$,~~~0.14$^f$ & 0.83$^p$,~~~0.14$^f$ & 0.00$^{p,~f}$     & 0.00$^{p,~f}$  \\
 $\Lambda_b(6152)~\pi$        & 0.06$^f$  & 14.49$^p$,~~~0.03$^f$  & 1.08$^p$,~~~0.14$^f$  & 1.10$^p$,~~~0.15$^f$ &  $-$               & 0.00$^{p,~f,~h}$  \\
 $B~N$                        & 2.13$^p$  & 0.12$^p$,              & 8.09$^p$              & 11.63$^f$            & 0.20$^f$           & 7.50$^f$   \\
 $B^\ast~N$                   & 7.76$^p$  & 20.81$^p$              & 13.74$^p$,~~~14.54$^f$  & 36.11$^p$,~~~8.66$^f$& 1.12$^f$         & 0.73$^f$   \\
\midrule[1pt]
 Total width                  & 88.36     & 102.06                 & 107.54                & 118.94               & 29.52              & 36.89   \\
\bottomrule[1pt]\bottomrule[1pt]
\end{tabular*}
\end{table*}

Here we would like to stress the particularity of $\Lambda_b(5912)\pi$ and $\Lambda_b(5920)\pi$ channels to search the $D$-wave $|5/2^+\rangle_3$ and $|7/2^+\rangle_3$ $\Sigma_b$ states. Since the $\Lambda_b(5912)$ and $\Lambda_b(5920)$ are below the threshold of $\Sigma_b(5815)\pi$, they can only decay through the three-body decay channel, i.e., $\Lambda_b\pi\pi$. As illustrated in Fig.~\ref{Fig3}, the $\Lambda_b(5912)$ and $\Lambda_b(5920)$ states with long lifetimes can fly far away from the decay vertex of the $D$-wave $|5/2^+\rangle_3$ and $|7/2^+\rangle_3$ $\Sigma_b$ states. This kind of process is very helpful for the LHCb detector to distinguish two pions which are produced from the $P$-wave $\Lambda_b^0$ state and the third pion from the initial $D$-wave $\Sigma_b$ state.

\subsection{$1D$ $\Xi_b^\prime$ baryons}

Up to now, only the 1$S$ $\Xi_b^\prime$ baryons, $\Xi_b^\prime(5935)$ and $\Xi_b^\ast(5955)$, have been established by experiments~\cite{Tanabashi:2018oca}. The masses and decay widths of 1$P$ and 2$S$  $\Xi_b^\prime$ baryons were given in Ref.~\cite{Chen:2018orb}.

As shown in Table \ref{table5}, the six 1$D$ $\Xi_b^\prime$ baryons are expected to appear in the region 6.40$-$6.50 GeV. Many decay channels are allowed for these excited $\Xi_b^\prime$ baryons (see Table \ref{table7}). The $|5/2^+\rangle_3$ and $|7/2^+\rangle_3$ states among the six $D$-wave $\Xi_b^\prime$ baryons may have narrower decay widths while the other four states seem to be broad. This prediction is similar to the $D$-wave $\Sigma_b$ baryons.

According to the results in Table \ref{table7}, it is promising for experimentalists to find the $|5/2^+\rangle_3$ and $|7/2^+\rangle_3$ $\Xi^\prime_b$ states in the $\Lambda_b(5620)~K$ channel. We also notice that the decay modes of $B\Lambda$ and $B^*\Lambda$ may make a significant contribution to the decays of $|1/2^+\rangle_1$, $|3/2^+\rangle_1$, $|3/2^+\rangle_2$, and $|5/2^+\rangle_2$ $\Xi^\prime_b$ states. Experimentalists could try to search for these four broad $\Xi^\prime_b$ states in the $B\Lambda$ and $B^*\Lambda$ channels.

\begin{table*}[htbp]
\caption{The predicted partial and total widths of $D$-wave $\Xi^{\prime}_b$ baryons (in MeV). The superscript letters $s$, $p$, $\cdots$, and $h$ mean that the corresponding decays occur via $s$-, $p$-, $\cdots$, and $h$-waves, respectively. If the partial width is predicted to be smaller than 0.01 MeV, it is denoted as ``0.00.'' If a decay channel is forbidden, it is denoted by ``$\times$.'' If the mass of an initial state is below the threshold of a decay channel, the partial width is denoted by ``$-$.'' The masses of 1$P$, 2$S$, and 1$D$ $\Xi^{(\prime)}_b$ baryons are borrowed from our previous work~\cite{Chen:2018orb}.} \label{table7}
\renewcommand\arraystretch{1.3}
\begin{tabular*}{170mm}{l@{\extracolsep{\fill}}cccccc}
\toprule[1pt]\toprule[1pt]
Final states    &  $\Xi^{\prime}_b(6486)$  & $\Xi^{\prime}_b(6488)$ & $\Xi^{\prime}_b(6456)$ & $\Xi^{\prime}_b(6457)$ & $\Xi^{\prime}_b(6407)$ & $\Xi^{\prime}_b(6408)$ \\
\midrule[1pt]
 $\Lambda_b(5620)~K$        & 1.21$^p$  & 1.23$^p$               &   $\times$            &  $\times$            &    8.47$^f$       &   8.47$^f$      \\
 $\Sigma_b(5815)~K$         & 4.87$^p$  & 1.18$^p$               &  15.01$^p$            &  2.38$^f$            &    0.93$^f$       &   0.54$^f$      \\
 $\Sigma^\ast_b(5835)~K$    & 3.06$^p$  & 7.48$^p$               & 3.54$^p$,~~~3.39$^f$  & 21.17$^p$,~~~1.52$^f$ &   0.70$^f$       &   0.96$^f$      \\

 $\Xi_b(5795)~\pi$          & 0.32$^p$  & 0.32$^p$               &   $\times$            &  $\times$            &    1.67$^f$       &   1.67$^f$      \\
 $\Xi^{\prime}_b(5935)~\pi$ & 0.03$^p$  & 0.01$^p$               &   0.14$^p$            &  0.26$^f$            &    0.23$^f$       &   0.13$^f$      \\
 $\Xi^{\ast}_b(5955)~\pi$   & 0.02$^p$  & 0.05$^p$               & 0.04$^p$,~~~0.52$^f$  & 0.21$^p$,~~~0.23$^f$ &    0.31$^f$       &   0.42$^f$      \\

 $\Lambda_b(5912)~K$        & 0.79$^s$  & 1.11$^d$               &   1.94$^d$            &  0.87$^d$            &        $-$        &   0.00$^g$      \\
 $\Lambda_b(5920)~K$        & 2.24$^d$  & 1.75$^s$,~~~1.13$^d$   &   1.38$^d$            &  1.39$^d$            &        $-$        &         $-$         \\
 $\Xi_b(6096)~\pi$          & 5.23$^s$  & 0.10$^d$               &   3.59$^d$            &  1.61$^d$            &    5.53$^d$       &   0.22$^g$      \\
 $\Xi_b(6102)~\pi$          & 0.27$^d$  & 4.88$^s$,~~~0.13$^d$   &   3.39$^d$            &  5.32$^d$            & 1.49$^d$,~~~0.31$^g$ & 6.76$^d$,~~~0.13$^g$ \\

 $\Xi^{\prime}_b(6150)~\pi$ &  $\times$ &  $\times$              &   0.19$^d$            &  0.20$^d$            &    $\times$       &   $\times$        \\
 $\Xi^{\prime}_b(6134)~\pi$ & 3.10$^s$  & 0.32$^d$               &   0.36$^d$            &  0.17$^d$            &    0.02$^d$       &    0.00$^g$           \\
 $\Xi^{\prime}_b(6139)~\pi$ & 0.53$^d$  & 3.31$^s$,~~~0.28$^d$   &   0.31$^d$            &  0.50$^d$            & 0.00$^{d,~g}$     & 0.02$^d$,~~~0.00$^g$  \\
 $\Xi^{\prime}_b(6094)~\pi$ & 1.09$^d$  & 0.56$^d$               & 6.04$^s$,~~~0.10$^d$  & 0.03$^d$,~~~0.01$^g$ & 0.33$^d$,~~~0.00$^g$ & 0.04$^d$,~~~0.00$^g$   \\
 $\Xi^{\prime}_b(6098)~\pi$ & 0.69$^d$  & 1.24$^d$               & 0.04$^d$,~~~0.01$^g$  & 6.41$^s$,~~0.10$^d$,~~0.00$^g$  & 0.07$^d$,~~~0.00$^g$  & 0.32$^d$,~~~0.00$^g$  \\

 $\Xi_b(6225)~\pi$          & 0.88$^p$  & 0.94$^p$               &   $\times$            &  $\times$            &    0.00$^f$       &   0.00$^f$      \\
 $\Xi_b(6327)~\pi$          & 2.24$^p$  & 0.26$^p$,~~~0.00$^f$   &   $-$                 &   $-$                &   $-$             &   $-$                  \\
 $\Xi_b(6330)~\pi$          & 0.00$^f$  & 1.88$^p$,~~~0.00$^f$   &   $-$                 &   $-$                &  $-$              &   $-$                  \\
 $B~\Lambda$                & 20.46$^p$ & 1.24$^p$,              & 17.16$^p$             & 1.49$^f$             & 0.00$^f$          & 0.03$^f$   \\
 $B^\ast~\Lambda$           & 17.13$^p$ & 49.43$^p$              & 5.51$^p$,~~~0.07$^f$  & 15.60$^p$,~~~0.05$^f$&  $-$              &  $-$                 \\
\midrule[1pt]
 Total width                & 64.16     & 78.83                  & 62.73                 & 59.52                & 20.06             & 19.71   \\
\bottomrule[1pt]\bottomrule[1pt]
\end{tabular*}
\end{table*}

\subsection{$1D$ $\Omega_b$ baryons}

So far, only the $\Omega_b(6046)^-$ state in the $bss$ baryon sector has been established~\cite{Tanabashi:2018oca}. However, there are reasons to believe that the situation could be changed in the near future. The widths of the 2$S$ and 1$P$ $\Omega_b$ states which were predicted in Ref.~\cite{Chen:2018vuc} imply that these excited $\Omega_b$ states are usually narrow. This feature can help experimentalists distinguish the signals of these low-lying $bss$ baryons from the background.

With the same method and parameters as before, we also calculate the masses and decays of six $D$-wave $\Omega_b$ states. The results are presented in Tables \ref{table5} and \ref{table8}, respectively. It is interesting to note that the $D$-wave excited $\Omega_b$ baryons are also expected to be narrow. Experimentalists may try to search for the $|1/2^+\rangle_1$, $|3/2^+\rangle_1$, $|3/2^+\rangle_2$, and $|5/2^+\rangle_2$ $\Omega_b$ baryons in the $\Xi^\prime_b(5935)K$ and $\Xi^\ast_b(5955)K$ channels. For the $|5/2^+\rangle_3$ and $|7/2^+\rangle_3$ $\Omega_b$ states, experiments may detect them around the 6.54 GeV in the $\Xi_b(5795)K$ channel.

\begin{table}[htbp]
\caption{The predicted widths of 1$D$ $\Omega_b$ baryons (in MeV).}\label{table8}
\renewcommand\arraystretch{1.3}
\begin{tabular*}{86mm}{@{\extracolsep{\fill}}ccccc}
\toprule[1pt]\toprule[1pt]
 1$D$ $\Omega_b$ states & $\Xi_b(5795)K$ &   $\Xi^\prime_b(5935)K$  & $\Xi^\ast_b(5955)K$ & Total  \\
\midrule[0.8pt]
 $\Omega_b(6599)$      & 0.10$^p$        &     0.65$^p$             &         0.41$^p$    & 1.16    \\
 $\Omega_b(6602)$      & 0.11$^p$        &     0.15$^p$             &         0.99$^p$    & 1.25    \\
 $\Omega_b(6578)$      &  $\times$       &     1.90$^p$             &  0.45$^p$,~~~0.53$^f$    & 2.88    \\
 $\Omega_b(6580)$      &  $\times$       &     0.41$^f$             &  2.68$^p$,~~~0.24$^f$    & 3.33    \\
 $\Omega_b(6542)$      & 3.32$^f$        &     0.24$^f$             &         0.19$^f$    & 3.75    \\
 $\Omega_b(6543)$      & 3.33$^f$        &     0.14$^f$             &         0.26$^f$    & 3.73    \\
\bottomrule[1pt]\bottomrule[1pt]
\end{tabular*}
\end{table}

\section{Discussions and conclusions}\label{sec5}

 There is not doubt that the LHCb Collaboration has played an important role in the research of the excited bottom baryons over the past years. Last year, two candidates for the $P$-wave bottom baryons, i.e., $\Xi_b(6227)^-$~\cite{Aaij:2018yqz} and $\Sigma_b(6097)^\pm$~\cite{Aaij:2018tnn}, were discovered by the LHCb. In particular, the $\Xi_b(6227)^-$ was the first excited bottom baryon which was identified by the OZI-allowed decay modes. So the observation of $\Xi_b(6227)^-$ could be be treated as a starting point for constructing the highly excited bottom baryon spectroscopy~\cite{Chen:2018orb}.

Just as expected, the LHCb Collaboration recently made a breakthrough in the research of the higher excited $\Lambda_b^0$ states~\cite{Aaij:2019amv}. Two almost degenerate narrow states, the $\Lambda_b(6146)^0$ and $\Lambda_b(6152)^0$, were observed in the $\Lambda_b^0\pi^+\pi^-$ spectrum. By comparing the excited energies of $\Lambda_b(6146)^0$ and $\Lambda_b(6152)^0$ to the corresponding $\Lambda_c^+$ states, we have found that they could be assigned as the 1$D$ $\Lambda_b^0$ states with $J^P=3/2^+$ and $J^P=5/2^+$, respectively.

The masses of the 1$D$ $\Lambda_b^0$ states predicted in Refs.~\cite{Capstick:1986bm,Chen:2014nyo,Chen:2018vuc} and the further investigation of the total decay widths also support the 1$D$ assignment of $\Lambda_b(6146)^0$ and $\Lambda_b(6152)^0$. However, the $\Sigma_b(5815)\pi$ is a main decay mode for the $\Lambda_b(6146)^0$ state if it is a $J^P=3/2^+$ state. This theoretical result is in disagreement with the measurement since no significant $\Lambda_b(6146)^0\rightarrow\Sigma_b^\pm\pi^\mp$ signals were observed at LHCb. Obviously, more experimental and theoretical efforts are desirable for the $\Lambda_b(6146)^0$ and $\Lambda_b(6152)^0$ in the future.

With the upcoming LHCb Upgrade I~\cite{Bediaga:2018lhg}, more excited bottom baryons will be discovered. In this work, we have also predicted the mass spectra and strong decay widths of other unseen 1$D$ excited bottom baryons. The following clues are suggested for searching for these unknown bottom baryons.

\begin{enumerate}[(1)]
\item We suggest to search for the $|5/2^+\rangle_3$ and $|7/2^+\rangle_3$ $\Sigma_b$ baryons in the $\Lambda_b(5620)~\pi$, $\Lambda_b(5912)~\pi$, and $\Lambda_b(5920)~\pi$ channels.

\item We suggest to search for the $|5/2^+\rangle_3$ and $|7/2^+\rangle_3$ $\Xi^\prime_b$ baryons in the $\Lambda_b(5620)~K$ channel.

\item Due to the very narrow decay widths and the simple decay modes, the $D$-wave $\Xi_b$ and $\Omega_b$ baryons are the potential states which could be found by experiments in the future.
\end{enumerate}


Obviously, the heavy baryon family including the charm and bottom baryons is being established step by step.  This interesting area of research deserves more attention from  theorists and experimentalists.

\section*{Acknowledgement}

X.L. is supported by the China National Funds for Distinguished Young Scientists under Grant No. 11825503 and the National Program for Support of Top-notch Young Professionals. B.C. is partly supported by the National Natural Science Foundation of China under Grants No. 11305003 and No. 11647301.


\end{document}